# Isotropic Metamaterial Stiffness Beyond Hashin-Shtrikman Upper Bound

Manish Kumar Singh[1], Chang Quan Lai[1,2*]

[1]School of Mechanical & Aerospace Engineering, Nanyang Technological University, 50 Nanyang Ave, Singapore, 639798, Singapore
[2]School of Materials Science & Engineering, Nanyang Technological University, 50 Nanyang Ave, Singapore, 639798, Singapore

[*] cqlai@ntu.edu.sg

## Abstract

Since its introduction more than 60 years ago, the Hashin-Shtrikman upper bound has stood as the theoretical limit for the stiffness of isotropic composites and porous solids, acting as an important reference against which the moduli of heterogeneous structural materials are assessed. Here, we show through first-principles calculations, supported by finite element simulations, that the Hashin-Shtrikman upper bound can be exceeded by the *isotropic* elastic response of an anisotropic structure constructed from an anisotropic material. The material and structural anisotropies mutually reinforce each other to realize the overall isotropic response, without incurring the mass penalty faced by the hybridization of geometries with complementary anisotropies. 3 designs were investigated (plate BCC, plate FCC and plate SC) but only plate SC yielded a solution for the anisotropic properties of the material, which are remarkably similar to that of single crystal nickel and single crystal ferrite.

**Main text**

Lightweight materials with high specific stiffness have always been key to improvements in vehicular fuel efficiency, which has garnered renewed scrutiny amidst the urgent need to decarbonize the transport sector due to climate change concerns[1,2]. Materials with stochastic porosity (*e.g.* foams or partially sintered powders) typically offer an isotropic mechanical response at the expense of low relative moduli, as their microstructural topologies, which are inherent to the manufacturing methods, tend to favor the softer deformation modes of beam bending or node rotation[3,4,5]. To achieve better relative stiffness, the microstructure can be imposed with a deterministic design that constrains the deformation of the resultant metamaterial lattice to a stretching/ compressive mode[6,7,8]. However, as with naturally occurring atomic crystal lattices, these well-defined metamaterial structures tend to be anisotropic, which can be detrimental to applications requiring multi-directional loading, in addition to raising the difficulty of computing the mechanical response of the structures under load.

Therefore, there is a strong motivation to develop a metamaterial lattice geometry that can offer isotropic global mechanical response and a high specific stiffness, as close to the theoretical limit of the Hashin-Shtrikman (H-S) upper bound as possible[9,10]. One of the earliest studies on this topic introduced isotropic 3D truss structures which consist of a simple cubic truss (*i.e.* struts aligned in the <100> directions) with additional diagonal struts that lie in the <111> or <110> directions[11]. These structures were rediscovered 22 years later by Gurtner and Durand in 2014[12], and 2 independent groups eventually mapped out all the possible combinations of Body-Centered

Cubic (BCC), Simple Cubic (SC) and Face-Centered Cubic (FCC) truss geometries that exhibit isotropic stiffness, publishing at almost the same time[13,14].

However, the specific relative moduli of these isotropic truss structures, which are subjected to a theoretical limit of $\frac{\bar{E}}{E_1 \rho_R} = \frac{1}{6}$ for $\rho_R \rightarrow 0$, fall far short of the highest possible isotropic modulus demarcated by the H-S upper bound[13,14]. To overcome this limitation, Berger *et. al.* eschewed beam-based truss structures for metamaterial lattices utilizing plate elements[10], introducing a hybrid plate Simple Cubic (*p*SC) and plate Face-Centered Cubic (*p*FCC) unit cell (*a.k.a.* plate Octet), with a thickness ratio of $\frac{t_{pSC}}{t_{pFCC}} = \frac{8\sqrt{3}}{9}$. This was quickly followed by a report that derived an infinite number of combinations of *p*BCC, *p*SC and *p*FCC lattices that can exhibit isotropic stiffness close to the H-S upper bound in the low density limit, as long as they were hybridized in the volume ratio of $x : 0.4–0.25x : 0.6–0.75x$, where $0 \leq x \leq 0.8$ [15]. Most recently, it was found that *p*FCC and *p*BCC can be nested into the enclosed spaces of *p*SC to enhance the specific modulus beyond those of hybridized geometries, up to a mere 2.1 % away from the H-S upper bound at a relative density of 0.6, the closest approach ever reported[16].

A key insight underlying these reports is that isotropic designs require the combination of unit cells with complementary (*i.e.* opposite) anisotropy. For instance, SC cells tend to be stiffer in the <100> directions while BCC or FCC cells tend to be stiffer in the <111> directions. Therefore, there is a relatively large range of volume fractions where the SC geometry can fuse with BCC or FCC to give isotropic response, but this is not so for FCC and BCC geometries, which have similar anisotropy profiles for their moduli.

Importantly, almost all of the studies on isotropic structural metamaterial design to date assume that the constituent materials are isotropic. To the best of our knowledge, there is no study on isotropic plate lattices constructed from an anisotropic material yet. Therefore, in this study, we investigated the feasibility of generating an isotropic lattice response from an anisotropic material architected into geometries with complementary anisotropy profiles. The geometries studied are the plate Simple Cubic (*p*SC), plate Face Centered Cubic (*p*FCC) and plate Body Centered Cubic (*p*BCC) designs, which were previously combined to great effect, with many hybrid structures shown to exhibit moduli at the theoretical limit of the H-S upper bound[10,15,16]. Moreover, they have been subjected to a similar analysis previously[15], which can help provide validation for our derivation results.

We began the investigation by following the general approach by Tancogne-Dejean *et. al.*[15] and applied the constraints of a global (*i.e.* lattice metamaterial) isotropic modulus to the periodically homogenized stiffness matrix of a specific lattice geometry. These constraints will then yield the local (*i.e.* material) anisotropic modulus profile required to realize the global isotropy. The details of the derivation can be found in the Supplementary Information, and only a brief outline is provided here.

The constitutive equation for an orthotropic constituent material making up the plate elements in the lattice metamaterial can be written as[17]

$$\begin{bmatrix}\sigma_{11}\\\sigma_{22}\\\sigma_{33}\\\sigma_{23}\\\sigma_{31}\\\sigma_{12}\end{bmatrix}=\begin{bmatrix}\lambda+2\alpha_1+\beta_1+2\mu+4\mu_1 & \lambda+\alpha_1+\alpha_2+\beta_3 & \lambda+\alpha_1 & 0 & 0 & 0\\\lambda+\alpha_1+\alpha_2+\beta_3 & \lambda+2\alpha_2+\beta_2+2\mu+4\mu_2 & \lambda+\alpha_2 & 0 & 0 & 0\\\lambda+\alpha_1 & \lambda+\alpha_2 & \lambda+2\mu & 0 & 0 & 0\\0 & 0 & 0 & (\mu+\mu_2) & 0 & 0\\0 & 0 & 0 & 0 & (\mu+\mu_1) & 0\\0 & 0 & 0 & 0 & 0 & (\mu+\mu_1+\mu_2)\end{bmatrix}\begin{bmatrix}\epsilon_{11}\\\epsilon_{22}\\\epsilon_{33}\\2\epsilon_{23}\\2\epsilon_{31}\\2\epsilon_{12}\end{bmatrix} \quad (1)$$

where $\lambda$, $\mu$, $\alpha_1$, $\alpha_2$, $\mu_1$, $\mu_2$, $\beta_1$, $\beta_2$, and $\beta_3$ are 9 independent material constants. Their relationships to the Young's modulus, *E*, shear modulus, *G*, and Poisson's ratio, *v*, can be found in the Supplementary Information. Note that Eq. (1) is only valid when the basis vectors ***a*** and ***b*** are aligned to the X- and Y- axes in the plane of the plate.

The 3D lattice strain, $\underline{\underline{\epsilon}}$, is related to the 2D in-plane strain of a single plate element, $\underline{\underline{\epsilon}}_0^p$, as

$$\underline{\underline{\epsilon}}_0^p = \underline{\underline{T}}\,\underline{\underline{\epsilon}}\,\underline{\underline{T}} \tag{2}$$

where $\underline{\underline{T}}$ is in the form of a square matrix which projects the general 3D lattice strain, $\underline{\underline{\epsilon}}$, onto the 2D plane of the plate and is given by $\underline{\underline{T}} = A(A^T A)^{-1} A^T$. Note that the number of lines under the symbols indicate the order of the tensor. Here $A$ is the rectangular matrix formed by the basis vectors, $\underline{a}$ and $\underline{b}$ in the plane of the plate. If $\underline{n}^p$ is the unit normal vector of the plate, given by $\{u, v, w\}^T$, such that $u^2 + v^2 + w^2 = 1$, then $\underline{n}^p$ makes an orthonormal triad with the basis vectors, $\underline{a}$ and $\underline{b}$. Assuming plane stress condition for the plate loading and applying the constraint of $\alpha_1 = \alpha_2$, the 3D plate strain, $\underline{\underline{\epsilon}}^p$, can be written as

$$\underline{\underline{\epsilon}}^p = \underline{\underline{\epsilon}}_0^p + \epsilon_{nn}^p\,\underline{n}^p \otimes \underline{n}^p = \underline{\underline{\epsilon}}_0^p - \left\{\left(\frac{\lambda+\alpha_1}{\lambda+2\mu}\right) tr(\underline{\underline{\epsilon}}_0^p)\right\} \underline{n}^p \otimes \underline{n}^p \tag{3}$$

Using the constitutive relation for orthotropic material in Eq. (1), the local stress field of the plate, $\underline{\underline{\sigma}}^p$, can be related to the 3D lattice strain, $\underline{\underline{\epsilon}}$, by

$$\underline{\underline{\sigma}}^p = \underline{\underline{\underline{C}}}^{pp}\underline{\underline{\epsilon}}^p = \underline{\underline{\underline{C}}}^{p}\underline{\underline{\epsilon}} \tag{4}$$

The volume average of the local stress fields, $\underline{\underline{\sigma}}^p$, of all the plates in the unit cell will then give the global metamaterial lattice stress field,

$$\underline{\underline{\sigma}}^M = \frac{1}{V_{UC}} \int_{V_{UC}} \underline{\underline{\sigma}} \, dv = \sum_{p \in UC} v^p \underline{\underline{\sigma}}^p = \left( \sum_{p \in UC} v^p \underline{\underline{C}}^p \right) \underline{\underline{\epsilon}} = \underline{\underline{C}}^M \underline{\underline{\epsilon}}, \quad (5)$$

where $\underline{\underline{C}}^M$ is the macroscopic lattice stiffness tensor, $v^p$ is the fraction of volume in a unit cell occupied by a single plate and $V_{UC}$ is the nominal volume of the unit cell.

In the limit of small relative density ($\rho_R \to 0$), where the intersecting volume of the plates is small, the relative density of the metamaterial lattice can be expressed as

$$\rho_R = \frac{V_p}{V_{UC}} = \frac{\rho_{UC}}{\rho_p} = \sum_{p \in UC} v^p, \quad (1)$$

where $V_p$ refers to the total volume of all the plates in the unit cell, $\rho_p$ refers to the mass density of the constituent material and $\rho_{UC}$ refers to the nominal mass density of the unit cell.

For an isotropic material or structure, it is known that the stiffness matrix should be

$$\underline{\underline{C}}^{iso} = \begin{bmatrix} \lambda_{iso} + 2\mu_{iso} & \lambda_{iso} & \lambda_{iso} & 0 & 0 & 0 \\ \lambda_{iso} & \lambda_{iso} + 2\mu_{iso} & \lambda_{iso} & 0 & 0 & 0 \\ \lambda_{iso} & \lambda_{iso} & \lambda_{iso} + 2\mu_{iso} & 0 & 0 & 0 \\ 0 & 0 & 0 & \mu_{iso} & 0 & 0 \\ 0 & 0 & 0 & 0 & \mu_{iso} & 0 \\ 0 & 0 & 0 & 0 & 0 & \mu_{iso} \end{bmatrix} \quad (7)$$

and its corresponding terms as invariants of $\underline{\underline{C}}^M$ would be

$$C_I^{iso} = (9\lambda_{iso} + 6\mu_{iso}) \text{ and } C_{II}^{iso} = (3\lambda_{iso} + 12\mu_{iso}) \quad (8)$$

Equating the strain energy for uniform strains of the global lattice structure to that of a homogenized isotropic material would lead to the equivalence of $\underline{\underline{C}}^M$ and $\underline{\underline{C}}^{iso}$, as well as their invariants. Computing this would then yield the isotropic lattice modulus and Poisson's ratio, as well as the direction-dependent elastic modulus of the constituent material that is required to realize this isotropy.

Interestingly, applying the above analysis on *p*BCC and *p*FCC did not yield any viable orthotropic material elastic parameters that can confer isotropy to the macroscopic lattice geometries (Supplementary Information). If an isotropic constituent material was assumed instead (*i.e.* $\alpha_1 = \alpha_2 = \mu_1 = \mu_2 = \beta_1 = \beta_2 = \beta_3 = 0$), our analysis reduces to Tancogne-Dejean's case[15], where isotropy can be achieved in a hybrid unit cell containing *p*BCC, *p*SC and *p*FCC lattices in the volume ratio of $x : 0.4 - 0.25x : 0.6 - 0.75x$, where $0 \leqslant x \leqslant 0.8$.

However, a viable result can be obtained for the *p*SC lattice, which is composed of 3 orthogonal plates with unit normals of $\{u, v, w\}^T = \{1,0,0\}^T, \{0,1,0\}^T$ and $\{0,0,1\}^T$ and a plate volume fraction of $v_p = \rho_R/3$ in the low relative density limit. Under these conditions, the analysis above indicates that the *p*SC geometry can exhibit an isotropic modulus of $\bar{E}$ and Poisson's ratio of $\bar{v}$,

$$\bar{E} = \frac{2E_1 \rho_R (2 - v_{12})}{3(2 + v_{12})(1 - v_{12})} \tag{9}$$

$$\text{and } \bar{v} = \frac{v_{12}}{2 + v_{12}}, \tag{10}$$

if the following constraints are satisfied for the orthotropic properties of the constituent material,

$$E_1 = E_2 = \frac{2G_{12}(v_{12}+1)(v_{12}-1)}{(v_{12}-2)}, \ G_{23} = G_{31}, \text{and } v_{31} = v_{32}. \tag{11}$$

Plotting the directional profile of the required elastic modulus described by Eq. (11), it can be observed that the material modulus is highest in <111>, followed by <110>, then <100> (Fig. 1A). The $E_{min}/E_{max}$ ratio is 0.49 while the anisotropy, which we will define using the Zener ratio, $Z =$ $\frac{2\bar{G}(1+\bar{\nu})}{\bar{E}}$, is 2.43. This elastic modulus profile is highly similar to single crystal face-centered cubic Nickel (Ni), as well as body-centered cubic ferrite (α-Fe), which exhibit a $E_{min}/E_{max}$ ratio of 0.45 and 0.52 respectively and an anisotropy of 2.5 and 2.15 respectively[18]. From Fig. 1B and 1C, it can be observed that the elastic modulus profiles of these materials only deviate from that of the ideal orthogonal plate material by $0.04E_{max}$, which would likely lie within the limits of experimental error. This result indicates that the elastic profile derived in Eq. (11) is not purely hypothetical and can potentially be realized with the right material and manufacturing techniques.

Importantly, the elastic profile of the anisotropic plate element is complementary to the stiffness profile of *p*SC geometry, which exhibits the highest stiffness in <100> directions and the lowest in <111> (Fig. 1A). This need for complementary stiffness profiles to achieve overall lattice isotropy is well documented in previous studies[14,15], where different unit cell geometries with opposite anisotropies were combined to obtain an isotropic hybrid unit cell design. The key difference is that hybridizing complementary lattice designs incurs a volume cost, because the other lattice occupies additional space within the first lattice geometry. However, this cost can be avoided when the complementary stiffness profile comes from the anisotropic nature of the material.

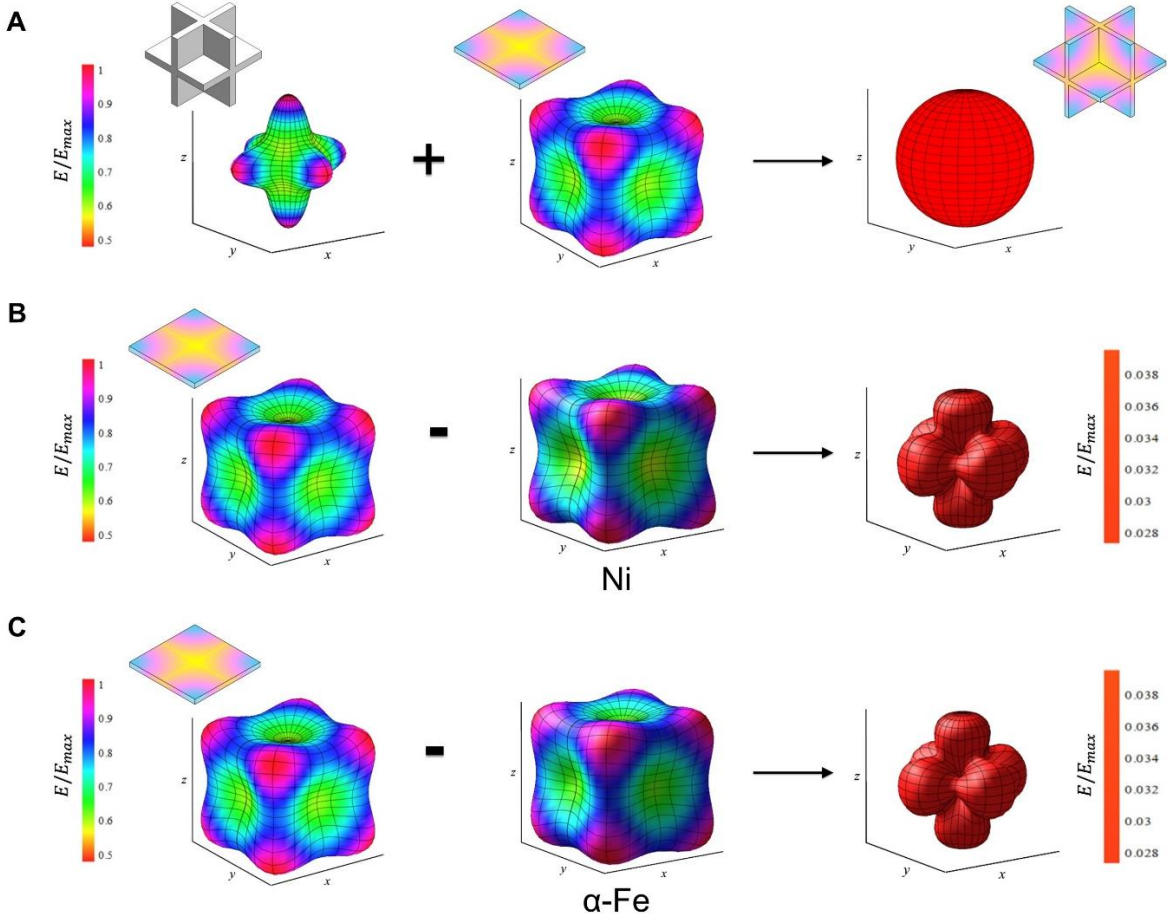

**Figure 1**: 3D surface plots showing the normalized Young's modulus (E/E$_{max}$) of different structures and materials, depicted in the insets, computed for different directions[19]. The gradient color plate represents the anisotropic plate, the properties of which were derived in Eq. (11). Note that Eq. (11) does not dictate any constraints for $E_3$, as lattice modulus only depends on the in-plane properties of the anisotropic plate, $E_1$ and $\nu_{12}$, according to Eq. (9) and Eq. (10). Therefore, we have assumed $E_3 = E_1$ here, since this is common amongst actual materials such as nickel (Ni) and ferrite (α-Fe). **(A)** Schematic illustration of how the combination of the Plate Simple Cubic structure and the anisotropic plate, which have complementary 3D surface plots, can lead to an isotropic Plate Simple Cubic (*p*SC) lattice. **(B)** Schematic illustration of the difference between 3D surface plots of the ideal anisotropic plate material and single crystal nickel (Ni). **(C)** Schematic illustration of the difference between 3D surface plots of the ideal anisotropic plate material and single crystal ferrite (α-Fe).

As a result, the specific relative stiffness, $\bar{E}/E_1\rho_R$, of the isotropic *p*SC lattice, for all possible Poisson's ratio (-1 ≤ $\nu_{12}$ ≤ 0.5) of the anisotropic constituent material, always exceeds that of the H-S upper bound, which represents the highest attainable stiffness for isotropic lattices constructed from isotropic constituent materials (Fig. 2A). Fixing $\nu_{12}$ yields a similar result – the isotropic $p$SC structure could attain a higher relative modulus, $\bar{E}/E_1$, than the H-S upper bound across the range of relative density investigated (Fig. 2B). This outcome was further confirmed by finite element analysis (FEA) conducted on the isotropic $p$SC metamaterial lattices in the <100>, <110> and <111> axes, which showed relative stiffnesses well above the H-S upper bound. The elastic moduli obtained through FEA agreed very well with the analytical trend for $\rho_R \leq 0.2$, but the simulation data remain reasonably close to the predicted values even up to $\rho_R = 0.5$, although our analysis was formulated in the limit of small $\rho_R$.

In addition to deviating from the analytical predictions at higher $\rho_R$, the elastic moduli of the isotropic $p$SC structure in different axes are also observed to diverge, which indicates a diminishing isotropy. A quantitative assessment using the Zener ratio, $Z = \frac{2\bar{G}(1+\bar{\nu})}{\bar{E}}$, confirms this result, as $Z$ increases from 0.98 at $\rho_R = 0.05$ to 1.20 at $\rho_R = 0.5$ (Fig. 2C), likely due to a breakdown of the plane stress and low relative density assumption used in our derivation, which affected the Poisson's ratio, $\bar{\nu}$, of the isotropic $p$SC lattice (Supplementary Information). While these results point to an isotropy best realized for $\rho_R \sim 0.1$, it is worth noting that $Z$ for the isotropic $p$SC structure remains within 10% of the ideal value of 1 for $\rho_R \leq 0.24$, which is still significantly better than anisotropic lattices such as $p$FCC and $p$SC constructed from isotropic materials[10] (Fig. 2C).

It is also worth noting that for low $\rho_R$, $\bar{E}/E_1$ of the isotropic $p$SC structure converges to that of anisotropic $p$SC lattice in its stiffest axis, the <100> direction[10] (Fig. 2B). This result suggests that the anisotropy of the plate material was enhancing the stiffness of the $p$SC lattice in the softer directions such as <110> and <111> to achieve isotropy, rather than reducing the lattice stiffness in the <100> directions, which is the case when geometries with complementary anisotropies are hybridized. For instance, by comparing $p$SC <100> and the hybridized isotropic SC-FCC lattices in Fig. 2B, it is clear that the relative stiffness of $p$SC in <100> directions fell after hybridizing with a plate FCC geometry. Again, this is because the hybridization incurs a volume cost that caused the data points of $p$SC to shift towards higher $\rho_R$ after hybridization. In contrast, by using the material's inherent anisotropy to shore up the soft directions of the $p$SC structure, there is no volume cost incurred and the data points of the $p$SC remained at the same $\rho_R$. This is a key reason that the specific stiffness attainable by the $p$SC structure via anisotropic plates is superior to that of previously reported isotropic lattices[10,15], which were composed of isotropic materials and hence, bounded by the Hashin-Shtrikman upper limit (Fig. 2B).

To summarize, we have shown, through analytical derivations from first principles, that isotropic metamaterial lattices can be obtained by arranging anisotropic plates in an anisotropic lattice geometry. Of the 3 geometries investigated, $p$SC, $p$BCC and $p$FCC, only $p$SC exhibited a solution for the material anisotropy, which were remarkably similar to the elastic modulus profile of single crystal nickel and single crystal ferrite. The relationships between the elastic parameters of the anisotropic plate and the properties of the isotropic $p$SC lattice were established analytically and it was used to show, together with results from finite element simulations, that the specific relative stiffness attainable by the isotropic $p$SC lattice is higher than the Hashin-Shtrikman upper limit, which previous isotropic lattices based on isotropic materials were bounded by. This study illuminates a new method of designing lightweight porous metamaterials with specific stiffness beyond the current theoretical limit, which could benefit the energy efficiency of vehicles for decarbonization of the transportation sector, if these structures can be manufactured.

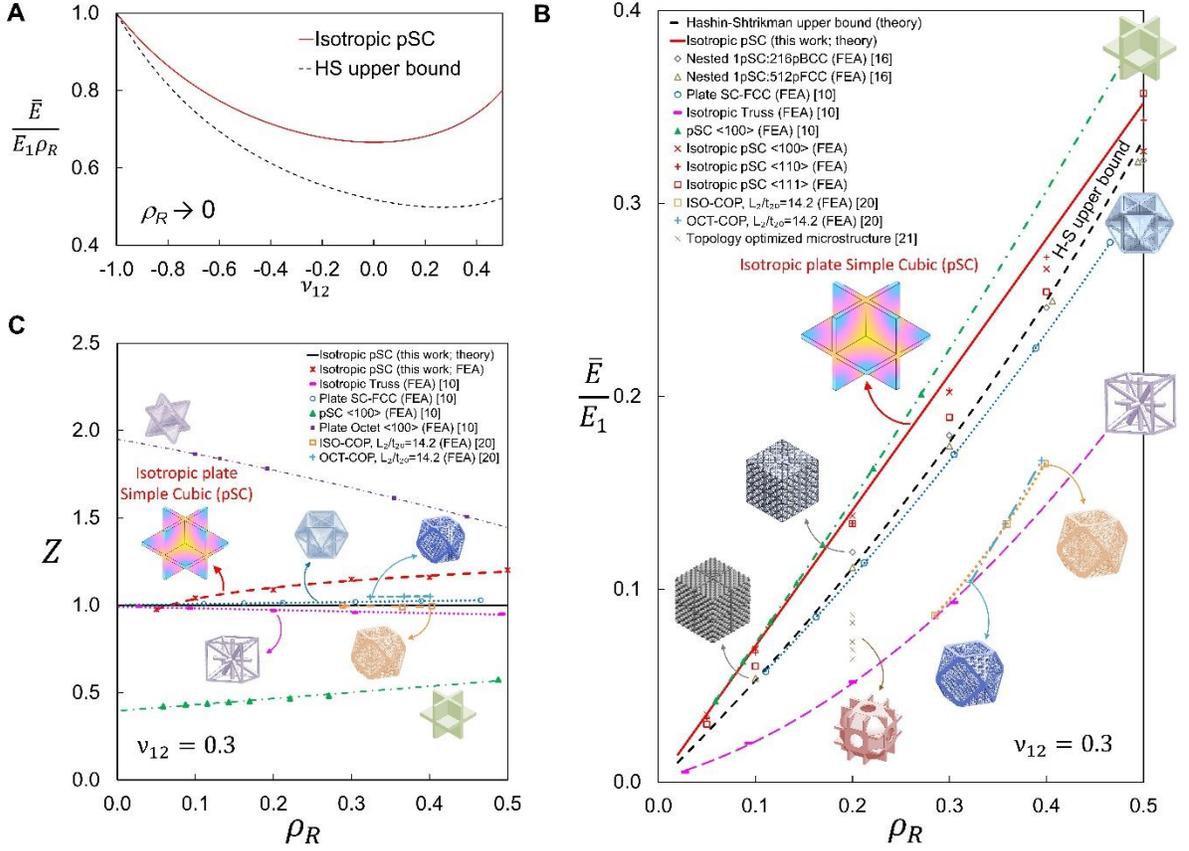

**Figure 2:** (**A**) Analytical results of the specific relative modulus, $\bar{E}/E_1\rho_R$, of the isotropic plate Simple Cubic lattice with respect to the Poisson's ratio, $\nu_{12}$, of the anisotropic plate, according to Eq. (9). The Hashin-Shtrikman upper bound (black dashed line) has been plotted as a reference. Plots of (**B**) relative modulus, $\bar{E}/E_1$, against relative density, $\rho_R$, and (**C**) Zener ratio, $Z = 2\bar{G}(1+\bar{\nu})/\bar{E}$, against relative density, $\rho_R$, for different lattice geometries composed of anisotropic material (gradient-colored plates) and isotropic material[10,16,20,21] for $\nu_{12}= 0.3$. The plate Octet and the plate Simple Cubic (*p*SC <100>) structures are anisotropic when constructed from an isotropic material, and have been included only for comparison. (insets) Schematic illustrations of the respective lattice structures.

## Methods

*Finite Element Simulations*

Finite Element (FE) simulations were carried out in Abaqus 2023 standard solver. Numerical simulations were performed on two types of structures: (a) a unit cell with periodic boundary conditions[22] and (b) 4×4×4 *p*SC unit lattices. For the first case, the homogenized effective elastic properties (*i.e.* Young's modulus, Poisson's ratio and shear modulus) of periodic representative unit cells were calculated for the <100> direction using an Abaqus plugin, EasyPBC[22]. The unit cell was discretized using hexahedral elements for the FE simulations. In the second case, a large 7×7×7 *p*SC lattice (Fig. S3A in Supplementary Information) was used to extract 4×4×4 unit cells with {100} and {110} faces (Fig. S3B) and {111} faces (Fig. S3C). Uniform displacement was applied normal to the faces and the resultant normal stress was used to compute the Young's modulus of the lattices.

## Supplementary Information

Additional details of the theoretical derivation and numerical simulations can be found in the supplementary information.

## Acknowledgements

The authors declare no conflicts of interest. This work was partially supported by C.Q.L's Nanyang Assistant Professorship grant (award no.: #022081-00001) and EDB-OSTIn grant (award no.: S22-19018-STDP).

## Data Availability

The data that supports the findings of this study are available within the article and its supplementary material.

Manish Kumar Singh[1], Chang Quan Lai[1,2*]

[1]*School of Mechanical & Aerospace Engineering, Nanyang Technological University, 50 Nanyang Ave, Singapore, 639798, Singapore*
[2]*School of Materials Science & Engineering, Nanyang Technological University, 50 Nanyang Ave, Singapore, 639798, Singapore*

[*] cqlai@ntu.edu.sg

# 1 First-principles derivation of anisotropic material properties required for an isotropic metamaterial lattice

Here, we derive the required constraints on elastic constants of the orthotropic constituent material for a global isotropic lattice. Vector and tensor quantities are denoted by bold letters with underlines, *e.g.*, a vector is denoted by $\underline{\boldsymbol{a}}$, a second order tensor by $\underline{\underline{\boldsymbol{A}}}$, and a fourth order tensor by $\underline{\underline{\underline{\boldsymbol{A}}}}$. Superscript T denotes the transpose of a vector or tensor, *e.g.*, $\underline{\underline{\boldsymbol{A}}}^{\boldsymbol{T}}$ denotes the transpose of the second order tensor $\underline{\underline{\boldsymbol{A}}}$. Cartesian coordinate system is used for the components of vector and tensor quantities. The scalar dot product of two vectors is denoted by $\underline{\boldsymbol{a}}.\underline{\boldsymbol{b}}$, and dyadic product of two vectors is denoted by $\underline{\boldsymbol{a}} \otimes \underline{\boldsymbol{b}}$, which is a second order tensor. Trace of a second order tensor (sum of the diagonal terms of the corresponding matrix) is denoted as $tr\left(\underline{\underline{\boldsymbol{A}}}\right)$. All constituent plates in the metamaterial lattice are assumed to have uniform thickness and made from the same orthotropic material. Linear elasticity and small deformations are assumed for the plates under load.

There are 9 independent elastic constants of the orthotropic material, denoted as $\lambda$, $\mu$, $\alpha_1$, $\alpha_2$, $\mu_1$, $\mu_2$, $\beta_1$, $\beta_2$ and $\beta_3$, with the following relationships to the Young's Modulus in three mutually orthogonal directions ($E_i$), Poisson's ratio ($\nu_{ij}$) and shear modulus ($G_{ij}$),

$$\lambda + 2\alpha_1 + \beta_1 + 2\mu + 4\mu_1 = \frac{1-\nu_{23}\nu_{32}}{E_2 E_3 \Delta}, \tag{S1}$$

$$\lambda + \alpha_1 + \alpha_2 + \beta_3 = \frac{\nu_{21}+\nu_{31}\nu_{23}}{E_2 E_3 \Delta}, \tag{S2}$$

$$\lambda + \alpha_1 = \frac{\nu_{31}+\nu_{21}\nu_{32}}{E_2 E_3 \Delta}, \tag{S3}$$

$$\lambda + 2\alpha_2 + \beta_2 + 2\mu + 4\mu_2 = \frac{1-\nu_{13}\nu_{31}}{E_1 E_3 \Delta}, \tag{S4}$$

$$\lambda + \alpha_2 = \frac{\nu_{32}+\nu_{31}\nu_{12}}{E_1 E_3 \Delta}, \tag{S5}$$

$$\lambda + 2\mu = \frac{1-\nu_{12}\nu_{21}}{E_1 E_2 \Delta}, \tag{S6}$$

$$\mu + \mu_2 = G_{23}, \ \mu + \mu_1 = G_{31}, \ \text{and} \ \mu + \mu_1 + \mu_2 = G_{12}. \tag{S7}$$

where $\Delta$ is given by the relation,

$$\Delta = \frac{1 - \nu_{12}\nu_{21} - \nu_{23}\nu_{32} - \nu_{13}\nu_{31} - 2\nu_{12}\nu_{23}\nu_{31}}{E_1 E_2 E_3}, \tag{S8}$$

and the Poisson's ratios are related by following equations,

$$\frac{\nu_{21}}{E_2} = \frac{\nu_{12}}{E_1}, \quad \frac{\nu_{31}}{E_3} = \frac{\nu_{13}}{E_1}, \ \text{and} \ \frac{\nu_{23}}{E_2} = \frac{\nu_{32}}{E_3}. \tag{S9}$$

The constitutive equation relating the stress tensor ($\underline{\underline{\sigma}}$) to the linear elastic strain tensor ($\underline{\underline{\epsilon}}$) is given by $\underline{\underline{\sigma}} = \underline{\underline{\underline{\underline{C}}}}\, \underline{\underline{\epsilon}}$, where $\underline{\underline{\underline{\underline{C}}}}$ is the fourth order stiffness tensor. If in an orthotropic material, the two families of fibres are orthogonal and along the unit basis vectors of the orthotropic plate plane $\underline{a}$ and $\underline{b}$, then the constitutive equation can be written as follows[1]:

$$\underline{\underline{\sigma}} = \{\lambda\, tr(\underline{\underline{\epsilon}}) + \alpha_1(\underline{a} \cdot \underline{\underline{\epsilon}}\, \underline{a}) + \alpha_2(\underline{b} \cdot \underline{\underline{\epsilon}}\, \underline{b})\}\underline{\underline{I}} + \{\alpha_1\, tr(\underline{\underline{\epsilon}}) + \beta_1(\underline{a} \cdot \underline{\underline{\epsilon}}\, \underline{a}) + \beta_3(\underline{b} \cdot \underline{\underline{\epsilon}}\, \underline{b})\}\underline{a}\otimes\underline{a} + \{\alpha_2\, tr(\underline{\underline{\epsilon}}) + \beta_3(\underline{a} \cdot \underline{\underline{\epsilon}}\, \underline{a}) + \beta_2(\underline{b} \cdot \underline{\underline{\epsilon}}\, \underline{b})\}\underline{b}\otimes\underline{b} + 2\mu\underline{\underline{\epsilon}} + 2\mu_1(\underline{a}\otimes\underline{a}\,\underline{\underline{\epsilon}} + \underline{\underline{\epsilon}}\,\underline{a}\otimes\underline{a}) + 2\mu_2(\underline{b}\otimes\underline{b}\,\underline{\underline{\epsilon}} + \underline{\underline{\epsilon}}\,\underline{b}\otimes\underline{b}) \quad (S20)$$

If $\underline{a}$ and $\underline{b}$ are chosen as vectors $\{1,0,0\}^T$ and $\{0,1,0\}^T$ respectively, then above equation can be written in Voigt notation (vector-matrix) form:

$$\begin{bmatrix}\sigma_{11}\\ \sigma_{22}\\ \sigma_{33}\\ \sigma_{23}\\ \sigma_{31}\\ \sigma_{12}\end{bmatrix} = \begin{bmatrix}\lambda + 2\alpha_1 + \beta_1 + 2\mu + 4\mu_1 & \lambda + \alpha_1 + \alpha_2 + \beta_3 & \lambda + \alpha_1 & 0 & 0 & 0\\ \lambda + \alpha_1 + \alpha_2 + \beta_3 & \lambda + 2\alpha_2 + \beta_2 + 2\mu + 4\mu_2 & \lambda + \alpha_2 & 0 & 0 & 0\\ \lambda + \alpha_1 & \lambda + \alpha_2 & \lambda + 2\mu & 0 & 0 & 0\\ 0 & 0 & 0 & (\mu + \mu_2) & 0 & 0\\ 0 & 0 & 0 & 0 & (\mu + \mu_1) & 0\\ 0 & 0 & 0 & 0 & 0 & (\mu + \mu_1 + \mu_2)\end{bmatrix}\begin{bmatrix}\epsilon_{11}\\ \epsilon_{22}\\ \epsilon_{33}\\ 2\epsilon_{23}\\ 2\epsilon_{31}\\ 2\epsilon_{12}\end{bmatrix} \quad (S11)$$

The general 3D strain for the lattice, $\underline{\underline{\epsilon}}$, is related to the 2D in-plane strain of a single plate element, $\underline{\underline{\epsilon}}_0^p$, as

$$\underline{\underline{\epsilon}}_0^p = \underline{\underline{T}}\, \underline{\underline{\epsilon}}\, \underline{\underline{T}} \quad (S12)$$

where $\underline{\underline{T}}$ is in the form of square matrix which projects the general 3D lattice strain, $\underline{\underline{\epsilon}}$, onto the 2D plane of the plate and is given by $\underline{\underline{T}} = A(A^T A)^{-1} A^T$. Here $A$ is the rectangular matrix formed by the basis vectors in the plane of the plate. If $\underline{n}^p$ is the unit normal vector of the plate given by $\{u, v, w\}^T$ such that $u^2 + v^2 + w^2 = 1$, then it makes an orthonormal triad with unit basis vectors $\underline{a}$ and $\underline{b}$ in the plane of the plate. The vectors $\underline{a}$, $\underline{b}$ and matrix $A$ are given as follows,

$$\underline{a} = \begin{bmatrix} 1 - \dfrac{u^2}{1+w} \\ \dfrac{-u\,v}{1+w} \\ -u \end{bmatrix}, \underline{b} = \begin{bmatrix} \dfrac{-u\,v}{1+w} \\ 1 - \dfrac{v^2}{1+w} \\ -v \end{bmatrix}, \underline{\underline{A}} = \begin{bmatrix} 1 - \dfrac{u^2}{1+w} & \dfrac{-u\,v}{1+w} \\ \dfrac{-u\,v}{1+w} & 1 - \dfrac{v^2}{1+w} \\ -u & -v \end{bmatrix}. \tag{S13}$$

Strain component along the normal of the plate ($\epsilon_{nn}^p$) is calculated by assuming a plane stress loading condition,

$$\sigma_{nn}^p = \lambda\, tr\left(\underline{\underline{\epsilon}}\right) + \alpha_1\, \epsilon_{aa}^p + \alpha_2\, \epsilon_{bb}^p + 2\mu\epsilon_{nn}^p = 0. \tag{S14}$$

Taking $\alpha_1 = \alpha_2$, and using the relation,

$$tr\left(\underline{\underline{\epsilon}}\right) = \epsilon_{aa}^p + \epsilon_{bb}^p + \epsilon_{nn}^p = tr\left(\underline{\underline{\epsilon_0^p}}\right) + \epsilon_{nn}^p, \tag{S15}$$

Eq. (S14) can be rearranged to give

$$\epsilon_{nn}^p = -\left(\dfrac{\lambda + \alpha_1}{\lambda + 2\mu}\right) tr\left(\underline{\underline{\epsilon_0^p}}\right). \tag{S16}$$

Now, the full 3D strain of the plate, $\underline{\underline{\epsilon}}^p$, can be written as

$$\underline{\underline{\epsilon}}^p = \underline{\underline{\epsilon_0^p}} + \epsilon_{nn}^p\, \underline{n}^p \otimes \underline{n}^p = \underline{\underline{\epsilon_0^p}} - \left\{\left(\dfrac{\lambda + \alpha_1}{\lambda + 2\mu}\right) tr\left(\underline{\underline{\epsilon_0^p}}\right)\right\} \underline{n}^p \otimes \underline{n}^p. \tag{S17}$$

Using constitutive relation for orthotropic material, the local stress field in the plane of the plate is given by the expression,

$$\begin{aligned}
\underline{\underline{\sigma}}^p &= \left\{\lambda tr\left(\underline{\underline{\epsilon}}^p\right) + \alpha_1\left(\underline{a}\cdot\underline{\underline{\epsilon}}^p\underline{a}\right) + \alpha_2\left(\underline{b}\cdot\underline{\underline{\epsilon}}^p\underline{b}\right)\right\}\underline{\underline{I}} \\
&+ \left\{\alpha_1 tr\left(\underline{\underline{\epsilon}}^p\right) + \beta_1\left(\underline{a}\cdot\underline{\underline{\epsilon}}^p\underline{a}\right) + \beta_3\left(\underline{b}\cdot\underline{\underline{\epsilon}}^p\underline{b}\right)\right\}\underline{a}\otimes\underline{a} \\
&+ \left\{\alpha_2 tr\left(\underline{\underline{\epsilon}}^p\right) + \beta_3\left(\underline{a}\cdot\underline{\underline{\epsilon}}^p\underline{a}\right) + \beta_2\left(\underline{b}\cdot\underline{\underline{\epsilon}}^p\underline{b}\right)\right\}\underline{b}\otimes\underline{b} + 2\mu\underline{\underline{\epsilon}}^p \\
&+ 2\mu_1\left(\underline{a}\otimes\underline{a}\underline{\underline{\epsilon}}^p + \underline{\underline{\epsilon}}^p\underline{a}\otimes\underline{a}\right) + 2\mu_2\left(\underline{b}\otimes\underline{b}\underline{\underline{\epsilon}}^p + \underline{\underline{\epsilon}}^p\underline{b}\otimes\underline{b}\right).
\end{aligned} \tag{S18}$$

Eq. (S18) can be simplified to the form, $\underline{\underline{\sigma}}^p = \underline{\underline{\underline{C}}}^{pp}\underline{\underline{\epsilon}}^p = \underline{\underline{\underline{C}}}^p\underline{\underline{\epsilon}}$, and rewritten in a vector-matrix form,

$$\begin{bmatrix}\sigma_{11}^p\\\sigma_{22}^p\\\sigma_{33}^p\\\sigma_{23}^p\\\sigma_{31}^p\\\sigma_{12}^p\end{bmatrix} = \begin{bmatrix}C_{1111}^p & C_{1122}^p & C_{1133}^p & C_{1123}^p & C_{1131}^p & C_{1112}^p\\ & C_{2222}^p & C_{2233}^p & C_{2223}^p & C_{2231}^p & C_{2212}^p\\ & & C_{3333}^p & C_{3323}^p & C_{3331}^p & C_{3312}^p\\ & & & C_{2323}^p & C_{2331}^p & C_{2312}^p\\ & sym & & & C_{3131}^p & C_{3112}^p\\ & & & & & C_{1212}^p\end{bmatrix}\begin{bmatrix}\epsilon_{11}\\\epsilon_{22}\\\epsilon_{33}\\2\epsilon_{23}\\2\epsilon_{31}\\2\epsilon_{12}\end{bmatrix}. \quad (S19)$$

The components in the modified plate stiffness tensor, $\underline{\underline{\underline{C}}}^p$, cannot be succinctly expressed in this document but can be found in the Maple file in the Supplementary Information. $\underline{\underline{\underline{C}}}^p$ has two invariants which do not change with plate orientation and are given by the expressions[2],

$$C_I^p = C_{1111}^p + C_{2222}^p + C_{3333}^p + 2(C_{1122}^p + C_{1133}^p + C_{2233}^p), \quad (S20)$$

and $C_{II}^p = C_{1111}^p + C_{2222}^p + C_{3333}^p + 2(C_{1212}^p + C_{3131}^p + C_{2323}^p), \quad (S21)$

which can be computed as

$$C_I^p = \beta_1 + \beta_2 + 2\beta_3 + 4\mu_1 + 4\mu_2 + \frac{8\mu^2 + 12\lambda\mu + 16\alpha_1\mu - 4\alpha_1^2}{(\lambda + 2\mu)}, \quad (S22)$$

and $C_{II}^p = \beta_1 + \beta_2 + 6\mu_1 + 6\mu_2 + \frac{12\mu^2 + 10\lambda\mu + 8\alpha_1\mu - 2\alpha_1^2}{(\lambda + 2\mu)}. \quad (S23)$

The macroscopic lattice stress field, $\underline{\underline{\sigma}}^M$, is given by the volume average of the local stress fields $\underline{\underline{\sigma}}^p$ of all the plates in the unit cell, so that

$$\underline{\underline{\sigma}}^M = \frac{1}{V_{UC}}\int_{V_{UC}}\underline{\underline{\sigma}}\,dv = \sum_{p\in UC}v^p\underline{\underline{\sigma}}^p = \left(\sum_{p\in UC}v^p\underline{\underline{\underline{C}}}^p\right)\underline{\underline{\epsilon}} = \underline{\underline{\underline{C}}}^M\underline{\underline{\epsilon}}, \quad (S24)$$

where $v^p$ is the fraction of volume in a unit cell occupied by a single plate, $V_{UC}$ is the nominal volume of the unit cell and $\underline{\underline{\boldsymbol{C}^M}}$ is the macroscopic lattice stiffness tensor given by matrix

$$\boldsymbol{C}^M = \begin{bmatrix} C^M_{1111} & C^M_{1122} & C^M_{1133} & C^M_{1123} & C^M_{1131} & C^M_{1112} \\ & C^M_{2222} & C^M_{2233} & C^M_{2223} & C^M_{2231} & C^M_{2212} \\ & & C^M_{3333} & C^M_{3323} & C^M_{3331} & C^M_{3312} \\ & & & C^M_{2323} & C^M_{2331} & C^M_{2312} \\ & sym & & & C^M_{3131} & C^M_{3112} \\ & & & & & C^M_{1212} \end{bmatrix}. \quad (S25)$$

The relative density, $\rho_R$, of the metamaterial lattice is given by,

$$\rho_R = \frac{V_p}{V_{UC}} = \frac{\rho_{UC}}{\rho_p} = \sum_{p \in UC} v^p, \quad (S26)$$

where $V_p$ refers to the total volume of all the plates in the unit cell, $\rho_p$ refers to the mass density of the constituent material and $\rho_{UC}$ refers to the mass density of the unit cell.

For an isotropic material or structure, it is known that the stiffness matrix should be

$$\boldsymbol{C}^{iso} = \begin{bmatrix} \lambda_{iso} + 2\mu_{iso} & \lambda_{iso} & \lambda_{iso} & 0 & 0 & 0 \\ \lambda_{iso} & \lambda_{iso} + 2\mu_{iso} & \lambda_{iso} & 0 & 0 & 0 \\ \lambda_{iso} & \lambda_{iso} & \lambda_{iso} + 2\mu_{iso} & 0 & 0 & 0 \\ 0 & 0 & 0 & \mu_{iso} & 0 & 0 \\ 0 & 0 & 0 & 0 & \mu_{iso} & 0 \\ 0 & 0 & 0 & 0 & 0 & \mu_{iso} \end{bmatrix} \quad (S27)$$

and its corresponding terms as invariants of $\underline{\underline{\boldsymbol{C}^M}}$ (Eq. (S20) and (S21)) would be

$$C_I{}^{iso} = (9\lambda_{iso} + 6\mu_{iso}) \text{ and } C_{II}{}^{iso} = (3\lambda_{iso} + 12\mu_{iso}) \quad (S28)$$

Equating the strain energy for uniform strains of the global lattice structure to that of a homogenized isotropic material would lead to the equivalence of all the components and invariants for $\underline{\underline{C}}^M$ and $\underline{\underline{C}}^{iso}$, which would then yield the isotropic lattice modulus and Poisson's ratio, as well as the anisotropic constituent material properties required to realize this isotropy.

## 1.1 Plate Simple Cubic (*p*SC) Geometry

The plate Simple Cubic (*p*SC) lattice is composed of 3 orthogonal plates with unit normals of $\{u, v, w\}^T = \{1,0,0\}^T, \{0,1,0\}^T$ and $\{0,0,1\}^T$ and each plate occupies a volume fraction of $v_p = \rho_R/3$ in the low relative density limit. The nonzero components of its lattice stiffness matrix are therefore,

$$C^M_{1111} = \frac{\rho_R}{3}\left(2\beta_1 + 8\mu_1 + \frac{8\mu^2 + 8\lambda\mu + 8\alpha_1\mu - 2\alpha_1^2}{(\lambda + 2\mu)}\right), \tag{S29}$$

$$C^M_{2222} = \frac{\rho_R}{3}\left(2\beta_2 + 8\mu_2 + \frac{8\mu^2 + 8\lambda\mu + 8\alpha_1\mu - 2\alpha_1^2}{(\lambda + 2\mu)}\right), \tag{S30}$$

$$C^M_{3333} = \frac{\rho_R}{3}\left(\beta_1 + \beta_2 + 4\mu_1 + 4\mu_2 + \frac{8\mu^2 + 8\lambda\mu + 8\alpha_1\mu - 2\alpha_1^2}{(\lambda + 2\mu)}\right), \tag{S31}$$

$$C^M_{1122} = C^M_{1133} = C^M_{2233} = \frac{\rho_R}{3}\left(\beta_3 + \frac{2\lambda\mu + 4\alpha_1\mu - \alpha_1^2}{(\lambda + 2\mu)}\right), \tag{S32}$$

$$C^M_{1212} = C^M_{2323} = C^M_{1313} = \frac{\rho_R}{3}(\mu + \mu_1 + \mu_2). \tag{S33}$$

All other components are zero and the two invariants of the tensor $\underline{\underline{C}}^M$ for *p*SC are

$$C_I^M = \rho_R \left( \beta_1 + \beta_2 + 2\beta_3 + 4\mu_1 + 4\mu_2 + \frac{8\mu^2 + 12\lambda\mu + 16\alpha_1\mu - 4\alpha_1^2}{(\lambda + 2\mu)} \right), \tag{S34}$$

$$C_{II}^M = \rho_R \left( \beta_1 + \beta_2 + 6\mu_1 + 6\mu_2 + \frac{12\mu^2 + 10\lambda\mu + 8\alpha_1\mu - 2\alpha_1^2}{(\lambda + 2\mu)} \right). \tag{S35}$$

Eq. (S29) – (S35) can be reduced to Tancogne-Dejean *et. al.*'s results[2] for *p*SC composed of isotropic constituent material if $\alpha_1 = \alpha_2 = \mu_1 = \mu_2 = \beta_1 = \beta_2 = \beta_3 = 0$, which serves as a useful check for the validity of the present derivation.

Equating the components of $\underline{\underline{C}}^M$ with $\underline{\underline{C}}^{iso}$ to get the material parameter constraints needed for an isotropic *p*SC lattice, we obtain

$$\beta_1 = \beta_2 \tag{S36}$$

$$\mu_1 = \mu_2 \tag{S37}$$

$$\frac{\rho_R}{3} \left( 2\beta_1 + 8\mu_1 + \frac{8\mu^2 + 8\lambda\mu + 8\alpha_1\mu - 2\alpha_1^2}{(\lambda + 2\mu)} \right) = \lambda_{iso} + 2\mu_{iso} \tag{S38}$$

$$\frac{\rho_R}{3} \left( \beta_3 + \frac{2\lambda\mu + 4\alpha_1\mu - \alpha_1^2}{(\lambda + 2\mu)} \right) - \frac{\rho_R}{3} (\mu + \mu_1 + \mu_2) = \lambda_{iso} - \mu_{iso} \tag{S39}$$

Similarly, equating the invariants of $\underline{\underline{C}}^M$ with corresponding terms of $\underline{\underline{C}}^{iso}$ yields

$$\lambda_{iso} = \frac{\rho_R}{15} \left( \beta_1 + \beta_2 + 4\beta_3 + 2\mu_1 + 2\mu_2 + \frac{4\mu^2 + 14\lambda\mu + 24\alpha_1\mu - 6\alpha_1^2}{(\lambda + 2\mu)} \right) \tag{S40}$$

$$\mu_{iso} = \frac{\rho_R}{15} \left( \beta_1 + \beta_2 - \beta_3 + 7\mu_1 + 7\mu_2 + \frac{14\mu^2 + 9\lambda\mu + 4\alpha_1\mu - \alpha_1^2}{(\lambda + 2\mu)} \right) \tag{S41}$$

Therefore, $\beta_3$ can be found by using Eq. (S40) and (S41) in Eq. (S39), giving

$$\beta_3 = 2\beta_1 + 4\mu_1 + \frac{4\mu^2 + 4\lambda\mu + 4\alpha_1\mu - \alpha_1^2}{(\lambda + 2\mu)} \tag{S42}$$

These constraints along with earlier assumption $\alpha_1 = \alpha_2$, can be reformulated using Eq. (S1) – (S9), into expressions involving elastic constants $E_1, E_2, E_3, \nu_{12}, \nu_{32}, \nu_{31}, G_{12}, G_{23}, G_{31}$ so that

$$E_1 = E_2 = \frac{2G_{12}(\nu_{12}^2 - 1)}{(\nu_{12} - 2)}, \; G_{23} = G_{31}, \text{ and } \nu_{31} = \nu_{32}. \tag{S43}$$

From Eq. (S40) and Eq. (S41), the isotropic lattice constants $\lambda_{iso}$ and $\mu_{iso}$ can be converted into the modulus, $\bar{E}$, and Poisson's ratio, $\bar{\nu}$, for the isotropic $p$SC lattice,

$$\bar{E} = \frac{2E_1\rho_R(2-\nu_{12})}{3(2+\nu_{12})(1-\nu_{12})} \text{ and } \bar{\nu} = \frac{\nu_{12}}{2+\nu_{12}} \tag{S44}$$

## 1.2 Plate Body Centered Cubic (*p*BCC) and Plate Face Centered Cubic (*p*FCC) Geometries

According to the present analysis, *p*FCC, *p*BCC or hybridization of either lattice with *p*SC unit cells cannot achieve isotropic lattice elastic response using orthotropic plates. The reason is that the first three diagonal terms of their stiffness matrices, $\underline{\underline{C^M}}$, are not equal and contain $\beta_3$, which prevents it from being independently isolated like the case with *p*SC lattice. Hybridization of *p*FCC, and *p*BCC with *p*SC unit cells can give isotropic lattice structures only when constituent material is isotropic (*i.e.* $\alpha_1 = \alpha_2 = \mu_1 = \mu_2 = \beta_1 = \beta_2 = \beta_3 = 0$), as discussed by Dejean et al.[2]

The first three diagonal terms for *p*BCC lattice are given by

$$\begin{aligned} C^M_{1111} &= \frac{\rho_{BCC}}{48}\Big(21\beta_1 + \beta_2 + 2\beta_3 \\ &+ \frac{8\lambda(12\mu + 11\mu_1 + \mu_2) + 16\mu(6\mu + 11\mu_1 + \mu_2) + 96\alpha_1\mu - 24\alpha_1^2}{(\lambda + 2\mu)}\Big), \end{aligned} \tag{S45}$$

$$C^M_{2222} = \frac{\rho_{BCC}}{48}\left(\beta_1 + 21\beta_2 + 2\beta_3 + \frac{8\lambda(12\mu+\mu_1+11\mu_2)+16\mu(6\mu+\mu_1+11\mu_2)+96\alpha_1\mu-24\alpha_1^2}{(\lambda+2\mu)}\right), \quad \text{(S46)}$$

$$C^M_{3333} = \frac{\rho_{BCC}}{6}\Bigg(\beta_1 + \beta_2 + \beta_3 \\ + \frac{6\lambda(2\mu+\mu_1+\mu_2)+12\mu(\mu+\mu_1+\mu_2)+12\alpha_1\mu-3\alpha_1^2}{(\lambda+2\mu)}\Bigg). \quad \text{(S47)}$$

The first three diagonal terms for *p*FCC lattice are given by

$$C^M_{1111} = \frac{\rho_{FCC}}{36}\Bigg((4\sqrt{3}+21)\beta_1 + (35-4\sqrt{3})\beta_2 + 8\beta_3 + 32\sqrt{3}(\mu_1-\mu_2) \\ + \frac{128\lambda(2\mu+\mu_1+\mu_2)+256\mu(\mu+\mu_1+\mu_2)+256\alpha_1\mu-64\alpha_1^2}{(\lambda+2\mu)}\Bigg) \quad \text{(S48)}$$

$$C^M_{2222} = \frac{-\rho_{FCC}}{9}\Bigg((2\sqrt{3}-7)\beta_1 - (2\sqrt{3}+7)\beta_2 - 2\beta_3 + 8\sqrt{3}(\mu_1-\mu_2) \\ + \frac{-32\lambda(2\mu+\mu_1+\mu_2)-64\mu(6\mu+\mu_1+\mu_2)-64\alpha_1\mu+16\alpha_1^2}{(\lambda+2\mu)}\Bigg) \quad \text{(S49)}$$

$$C^M_{3333} = \frac{4\rho_{FCC}}{9}\Bigg(\beta_1 + \beta_2 + 2\beta_3 \\ + \frac{8\lambda(2\mu+\mu_1+\mu_2)+16\mu(\mu+\mu_1+\mu_2)+16\alpha_1\mu-4\alpha_1^2}{(\lambda+2\mu)}\Bigg) \quad \text{(S50)}$$

## 2 Finite element convergence analysis

Normalized Young's modulus of 4x4x4 lattices (relative density = 0.2) in <100> direction for different number of hexahedral elements has been plotted to show the mesh convergence of the finite element analysis (Fig. S1).

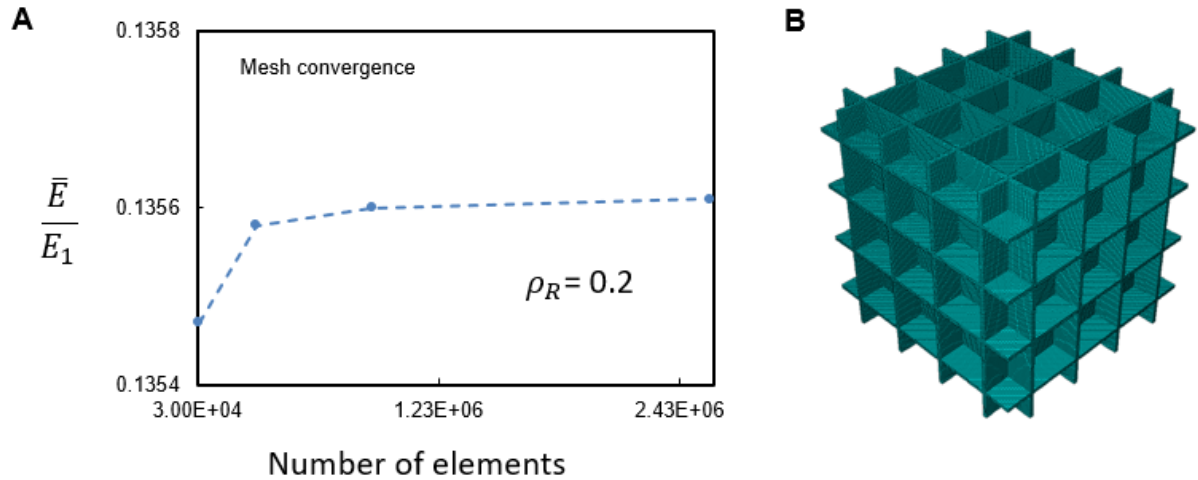

**Supplementary Figure S3**: **(A)** Mesh convergence plot and **(B)** solid element mesh of 4x4x4 lattices.

# 3 Comparison of theoretical and FEA results for effective Poisson's ratio

The analytical results for Poisson's ratio, $\bar{\nu}$, of isotropic $p$SC lattices (Eq. (S44)) and HS upper bound[3,4] in the low relative density limit are primarily dependent on the Poisson's ratio of the constituent material, $\nu_{12}$. The lattice Poisson's ratio of HS upper bound is the same as that for isotropic plate lattices when $\nu_{12} = -1$, but rises above $\bar{\nu}$ for isotropic $p$SC as $\nu_{12}$ increases (Fig. S2A). However, FEA results for isotropic $p$SC increasingly deviate from the analytical expectation with increasing relative density, likely due to a breakdown of the plane stress and low relative density assumptions used in our derivation (Fig. S2B). The deviation in Poisson's ratio, in turn, led to the deviation of Zener ratio observed in Fig. 2C of main text.

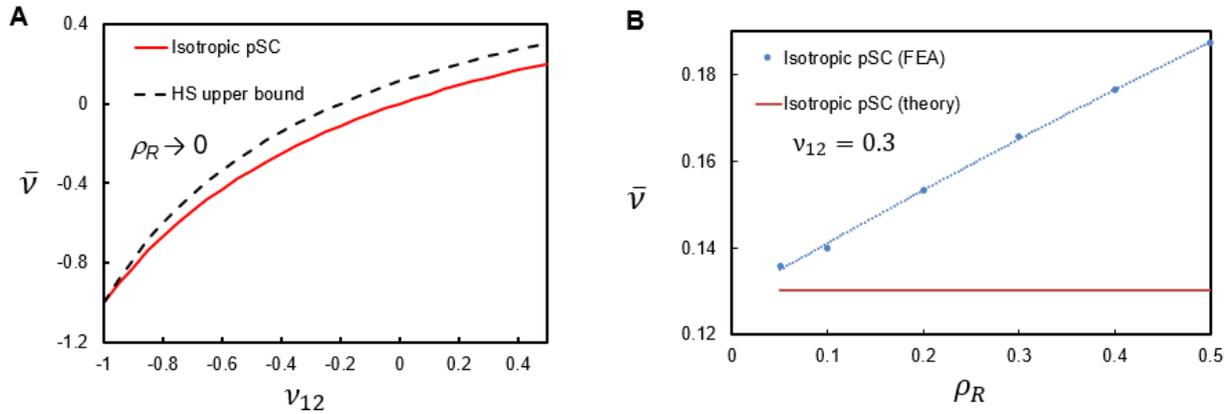

**Supplementary Figure S4:** **(A)** Effective Poisson's ratio, $\bar{\nu}$, for isotropic plate lattices and HS upper bound as a function of Poisson's ratio of constituent material, $\nu_{12}$. **(B)** Comparison of theoretical and FEA results for effective Poisson's ratio, $\bar{\nu}$, with relative density, $\rho_R$.

# 4 Models used for finite element simulations

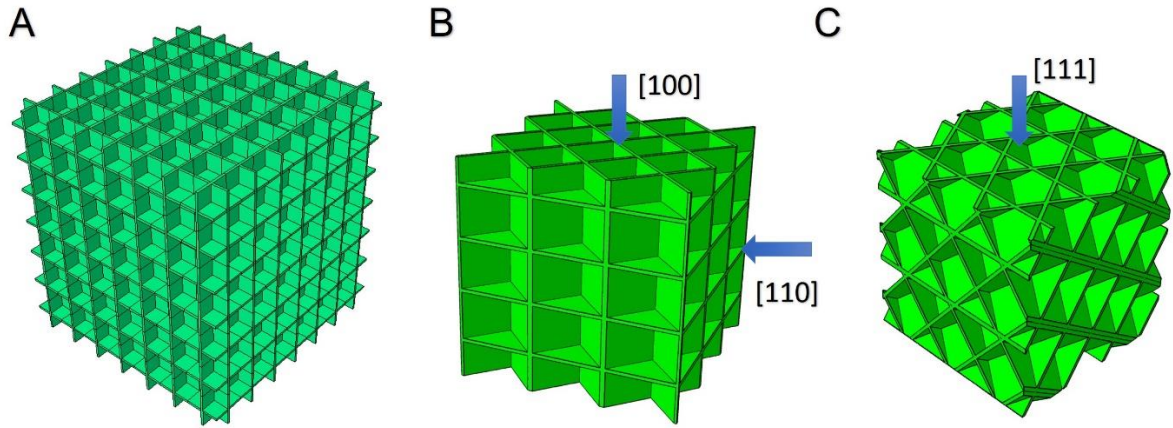

**Supplementary Figure S5**: **(A)** Plate SC lattice structure (7×7×7 unit cells) with {100} faces and 4×4×4 lattices with **(B)** {100} and {110} faces and **(C)** {111} faces, that were extracted from the model in **(A)**.